\documentclass[numberedappendix]{emulateapj}

\begin{document}

\title{AGN Feedback and Cooling Flows: Problems with simple
  hydrodynamical models}
\author{John C. Vernaleo and Christopher S. Reynolds}
\affil{Department of Astronomy, University of Maryland, College Park,
  MD 20742}
\email{vernaleo@astro.umd.edu, chris@astro.umd.edu}

\begin{abstract}
In recent years it has become increasingly clear that Active Galactic
Nuclei, and radio-galaxies in particular, have an impact on large
scale structure and galaxy formation.  In principle, radio-galaxies
are energetic enough to halt the cooling of the virialized
intracluster medium (ICM) in the inner regions of galaxy clusters,
solving the cooling flow problem and explaining the high-mass
truncation of the galaxy luminosity function.  We explore this process
through a series of high resolution, three dimensional hydrodynamic
simulations of jetted active galaxies that act in response to
cooling-mediated accretion of an ICM atmosphere.  We find that our
models are incapable of producing a long term balance of heating and
cooling; catastrophic cooling can be delayed by the jet action but
inevitably takes hold.  At the heart of the failure of these models is
the formation of a low density channel through which the jet can
freely flow, carrying its energy out of the cooling core.  It is
possible that this failure is due to an over-simplified treatment of
the fast jet (which may underestimate the ``dentist drill'' effect).
However, it seems likely that additional complexity (large-angle jet
precession or ICM turbulence) or additional physics
(magnetohydrodynamic effects and plasma transport processes) is
required to produce a spatial distribution of jet heating that can
prevent catastrophic cooling.  This work also underscores the
importance of including jet dynamics in any feedback model as opposed
to the isotropically inflated bubble approach taken in some previous
works.
\end{abstract}

\keywords{cooling flows --- galaxies: active --- galaxies: formation
  --- galaxies: jets --- hydrodynamics --- X-rays: galaxies: clusters}

\maketitle

\section{Introduction}

In recent years, it has become increasingly clear that Active Galactic
Nuclei (AGN) may well have an impact on large scale structure and
galaxy formation.  The accepted framework for galaxy formation states
that baryonic matter falls into developing dark matter halos and the
resulting accretion shocks raising it to the approximately the virial
temperature of the halo.  The ``cold'' baryonic components of the
galaxy then form via radiative cooling of this shocked gas.  In the
absence of any feedback processes (i.e., with gravitational collapse
followed by radiative cooling alone), the galaxy mass function would
have to essentially follow the dark matter halo mass function --- in
essence, the baryons that are within the turn-around radius of the
developing dark matter halo are trapped and fated to eventually form
the baryonic galaxy at the halo's center.  However, as has been long
known, this is clearly not the case.  At both the highest and lowest
masses, there is a deficit of galaxies compared with dark matter
halos.  These major discrepancies indicates that multiple,
non-gravitational, feedback mechanisms act on galaxy formation.

At the low luminosity/mass end of the galaxy distribution, the deficit
can be explained entirely by feedback from star
formation \citep{1974MNRAS.169..229L,1986ApJ...303...39D}.
Once the first (massive) stars to form from the cooling baryons
start to supernova, superwinds from the protogalaxy can expel a
significant fraction of the remaining baryons from the (shallow)
gravitational potential well of the dark matter halo.  The explanation
for the deficit at the high mass end is not so clear, though.
Inefficient cooling \citep{1977MNRAS.179..541R} of baryons that have
been shocked to the high virial temperature (exceeding $10^7\,{\rm K}$)
of large dark matter halos can account for some of the deficiency but
is insufficient.  An additional mechanism is required that is
substantially more efficient than star formation \citep[e.g., see
discussion in ][]{2003ApJ...599...38B}.  Many authors have suggested
that radio-loud AGN provide this additional feedback.

If radio-loud AGN do indeed regulate the formation of the most massive
galaxies, this is probably one of the few aspects of galaxy formation
that can be studied in detail in the local universe: i.e., in the
cores of cooling galaxy clusters.  Galaxy clusters possess the largest
dark matter halos found in the Universe and contain virial baryons
(the intracluster medium; ICM) at X-ray emitting temperatures.  The
ICM of galaxy clusters has been intensively studied by every X-ray
observatory since its discovery in the earliest days of X-ray
astronomy \citep{1966ApJ...146..955F}, but imaging spectroscopy by
{\it Chandra} and {\it XMM-Newton} have raised these studies to an
unprecedented level.  At its simplest, these observations provide
measurements of the density and temperature of the ICM, which allows a
radiative cooling time to be computed.  In the central regions of most
relaxed clusters, the cooling time is often significantly shorter than
a Hubble time (and often as low as ${\rm few}\times 10^8$ years).
With such short cooling times, there must be either a growing ``sink''
of cooled gas or a heat source acting on the cluster center.
Observationally, there are very strong limits of the amount of cool
gas present.  In particular, there is usually insufficient star
formation seen in the central cD galaxy to account for the cooling
flow [although there is clearly star formation
occurring; \citet{2004ApJ...612..131O,2005ApJ...635L...9H}].  The
detailed discrepancy between the X-ray measured cooling rate and the
lack of cooled gas forms the classic ``cooling flow
problem'' \citep{1994ARA&A..32..277F}.  Dispersive spectroscopy by
{\it XMM-Newton} deepened this mystery.  While most clusters clearly
show evidence for radiative cooling of the ICM from the virial
temperature $T_{\rm vir}$ to about $T_{\rm vir}/3$, the absence of
lower-ionization iron lines sets tight limits on the emission measure
of gas below $T_{\rm vir}/3$ (which would be expected to be a prolific
radiator; \citet{2001A&A...365L.104P,2001A&A...365L..87T}).

Of course, there is an obvious link between the apparent inability of
the ICM to cool below X-ray emitting temperatures and the deficit of
massive galaxies.  It is precisely the cD galaxies at the center of
rich galaxy clusters which are ``trying'' to grow above the observed
cutoff in the galaxy mass function via the radiative cooling of the
core regions of the ICM.  This connection strongly suggests that
solutions to the cooling flow problem that rely on ``hiding'' the
signatures of cooling (e.g., through strong metallicity
inhomogeneities or mixing layers) are doomed to fail, thereby
increasing the motivation for providing a viable heat source for the
cluster core.

Radio-loud AGN are very energetic and are frequently found in the
giant elliptical (cD) galaxies in the center of cooling
clusters \citep{1977ApJ...217...34B}, which makes them ideal
candidates for the heat source needed to offset cooling cluster.  Even
putting galaxy formation and cooling flow arguments aside, however, it
is clear that radio loud AGN have an impact on their environment.
First seen in {\it Einstein} and {\it ROSAT}
observations \citep{1987ApJ...312..101F, 1993MNRAS.264L..25B,
1995MNRAS.274L..67B, 1994MNRAS.270..173C,1998ApJ...501..126H}, {\it
Chandra} has studied numerous examples of ICM
bubbles \citep{2000MNRAS.318L..65F, 2000ApJ...534L.135M,
2001ApJ...558L..15B, 2002ApJ...579..560Y}, ghost
bubbles \citep{2001ApJ...562L.149M, 2002ApJ...569L..79H,
2004ApJ...606..185C, 2000MNRAS.318L..65F},
ripples \citep{2003MNRAS.344L..43F,2005MNRAS.360L..20F},
shells \citep{2000MNRAS.318L..65F}, and filaments that are clearly
associated with a central radio-loud AGN.  There is a large body of
numerical work which models various aspects of these interactions.
Early hydrodynamic models of the putative jet-cocoon/shock structure
in Cygnus-A were presented by \citep{1997MNRAS.284..981C}.  More
recent numerical investigations have explored the buoyant evolution of
the cocoon after its supersonic expansion
phase \citep{2001ApJ...554..261C,
2001MNRAS.325..676B,2002Natur.418..301B,
2002MNRAS.332..271R,2003MNRAS.339..353B,2004MNRAS.348.1105O,2004MNRAS.350L..13O,2004MNRAS.355..995D,2004ApJ...601..621R,2005A&A...429..399Z}, the
effect of plasma transport processes \citep{2004ApJ...615..675R,
2005MNRAS.357..242R} and the action of magnetic
fields \citep{2004ApJ...601..621R,2005ApJ...624..586J}.

Despite the reasons to look to AGN heating as the solution to the
cooling flow problem (and hence the regulator of high mass galaxy
formation), there is a problem with this picture.  Direct
observational signatures of ICM heating remain elusive in the
observations despite the existence of some very deep {\it Chandra} and
{\it XMM-Newton} observations of nearby vigorous ICM/AGN interactions.
Strong shocks are not seen in most systems even though they are seen
in many simulations that capture strong AGN heating (however, see
\citet{2004MNRAS.348.1105O} for a discussion of the difficulty of
seeing shocks in slow jet simulations).  Furthermore, the recent
sample of radio-galaxy blown ICM bubbles
by \citet{2004ApJ...607..800B} suggests that the total work done by
the AGN to inflate the observed ICM cavities may not be sufficient (by
approximately an order of magnitude) to offset cooling in many
clusters.  This may point to the importance of heating by the
dissipation of sound waves \citep{2005MNRAS.363..891F} or the decay of
global ICM modes \citep{2004MNRAS.348.1105O,2005MNRAS.357..381R}

In this paper we perform high-resolution, three dimensional ideal
hydrodynamic simulations of AGN feedback in a relaxed cooling cluster.
Unlike many of the previous 3-d investigations [with the notable
exceptions
of \citet{2003MNRAS.339..353B,2004MNRAS.348.1105O,2004MNRAS.350L..13O}],
we simulate the injection of a supersonic jet into the ICM atmosphere
rather than starting with an initial condition of a pre-inflated,
static radio cocoon.  We consider this an important issue --- only
through a direct modeling of the jet can we hope to be able to capture
the jet-induced ICM shock heating as well as the complex internal
dynamics of the cocoon.  While the effect of the jet on the ICM is
treated from first principles (through the evolution of the equations
of hydrodynamics), the enormous range of scales between the ICM core
and the gravitational sphere of influence of the black hole prevents
us from treating the formation of the jet or the AGN fueling properly.
In this paper, we introduce several different feedback prescriptions
by which the AGN jet power is related to the ICM properties in the
innermost region.  Our goal is to construct a model system in which
AGN heating and ICM cooling are, in the long term, in balance.  We
find that our ideal hydrodynamic models {\it fail} to achieve this
balance.  Assuming that the AGN feedback picture is indeed correct, we
conclude that our models must be failing to capture some crucial
aspect of the system.

In Section~\ref{sec:setup}, the basic setup for our simulations is
described, including a description of the model ICM atmosphere and our
hydrodynamic code.  Section~3 presents our results for the different
feedback scenarios considered.  We discuss our results along with
possible ways to rescue the AGN feedback hypothesis in
Section~\ref{sec:disc}.  Finally our modifications to the ZEUS-MP code
are explained in Appendix~\ref{append:zeus}.

\section{Basic Setup}
\label{sec:setup}

We aim to model the ICM of a cluster and its interaction with a
central radio galaxy.  Our basic picture is a spherical cluster
consisting initially of stationary dark matter and gas.  The gas is
initially in hydrostatic equilibrium, but is cooling through optically
thin thermal emission.  As the gas cools, some of it will flow across
the inner radial boundary of the simulation and is no longer
simulated.  The amount of gas to cross this boundary is used as our
primary diagnostic of the cooling flow.

Initially, the cluster is spherically symmetric and isothermal.  The
gas is setup with a $\beta$-model profile,
\begin{equation}
  \label{eq:beta}
  \rho(r)=\frac{1}{[1+(\frac{r}{r_0})^2]^{3/4}}.
\end{equation}
This is a simple analytic fit to cluster mass profiles.  The
gravitational potential,
\begin{equation}
  \label{eq:grav}
  \Phi=\frac{c^2_s}{\gamma}\ln(\rho),
\end{equation}
is the set so the initial gas configuration is in hydrostatic
equilibrium.  The potential is assumed to be generated by a static
distribution of dark matter which remains fixed throughout the
simulation. The self gravity of the gas is ignored (we note
that in a standard cosmology, the gas mass is only around $14\%$ of
the total cluster mass \citep{2005ApJ...634..964O}).

Spherical polar $(r,\theta,\phi)$ coordinates are used.  All
simulations were run on a $200\times200\times100$ cell grid, with
enhanced resolution near the center of the cluster and near the jet
axis.  This corresponds to a physical grid with a inner radius at
$r=10\,{\rm kpc}$ and an outer radius at $r=1000\,{\rm kpc}$.  The
angular coordinate was only allowed to vary from $\theta=0$ to
$\theta=\frac{\pi}{2}$.  This effectively only covers half the cluster
and therefore only allows for one jet.  Reflecting boundaries were use
to mimic the effect of the missing half of the cluster. This aids in
maintaining a high-resolution while keeping a reasonable number of
grid cells.

In ideal (adiabatic) hydrodynamics, the equations of hydrodynamics can
be written in dimensionless form and, hence, the results of a single
simulation may be scaled to a whole family of problems with different
mass, size and time scales (e.g., see \citet{2002MNRAS.332..271R}).
Since we must add radiative cooling to our simulations (see
Section~\ref{sec:cool}), we are forced into a set of physical units.
The values for the rich cluster of \citet{2002MNRAS.332..271R} were
used.  This gives us a core radius of $r_o=100\,{\rm kpc}$, a code
length unit of $r=50\,{\rm kpc}$, time units of $50\,{\rm Myrs}$, a
sound speed $c_s=1000\,{\rm km}\,{\rm s}^{-1}$.  The resulting total
radiative luminosity of the model ICM is $1.22\times10^{45}\,{\rm
erg}\,{\rm s}^{-1}$.

\subsection{Code}
\label{sec:code}

The simulations were performed using the ZEUS-MP code.  A modified
version of the original National Center for Supercomputing
Applications (NCSA) release was used.  We have made our modified
version publicly
available\footnote{http://www.astro.umd.edu/$\sim$vernaleo/zeusmp.html}.
A further discussion of this version of ZEUS-MP detailing our
modifications is given in Appendix~\ref{append:zeus}.

ZEUS-MP is a parallel version of the ZEUS magnetohydrodynamic
code \citep{1992ApJS...80..753S,1992ApJS...80..791S}.  ZEUS is a
fixed-grid, time-explicit Eulerian code which uses an artificial
viscosity to handle shocks.  When operated using van Leer advection,
as in this work, it is formally of second order spatial accuracy.  The
work reported here used this code in a pure hydrodynamic mode.  All
simulations were done on the University of Maryland, Astronomy
Department's GNU/Linux Beowulf cluster with run times of the order of
one to two months for four processor simulations.

ZEUS solves the standard hydrodynamic equations, which, including
radiative cooling (Section~\ref{sec:cool}) are
\begin{equation}
  \label{eq:hd1}
  \frac{D \rho}{D t}+\rho\nabla\cdot v = 0,
\end{equation}

\begin{equation}
  \label{eq:hd2}
  \rho\frac{D v}{D t}=-\nabla P -\rho\nabla\Phi,
\end{equation}

\begin{equation}
  \label{eq:hd3}
  \rho\frac{D }{D t} \left( \frac{e}{\rho} \right) =-P\nabla\cdot v - \Lambda,
\end{equation}
where
\begin{equation}
  \label{eq:hd4}
  \frac{D }{D t}\equiv\frac{\partial}{\partial t} + v\cdot\nabla.
\end{equation}
Stability is determined by the usual Courant-Friedrichs-Lewy (CFL)
condition.  As the jet is typically very supersonic, its speed
provides the limiting timestep during most of the simulations.

\section{Specific Models}
\label{sec:model}

A set of thirteen 3-d simulations were performed.  Table~\ref{t:sims}
lists the simulations along with some relevant parameters.  In the
following sections we will describe each simulation along with
presenting the results of each.

\begin{table}

\caption{List of Simulations}
\begin{center}
  \begin{tabular}{c c c c c c }\hline \hline
    Name & Feedback & Efficiency & Delay & Rotation \\ \hline
    Run A & none & NA & NA & NA\\ 
    Run B & strong jet & NA & NA & NA\\ 
    Run C & weak jet & NA & NA & NA\\ 
    Run D & simple feedback & 0.0001 & NA & NA\\ 
    Run E & simple feedback & 0.00001 & NA & NA\\ 
    Run F & delayed feedback & 0.0001 & short & NA\\ 
    Run G & delayed feedback & 0.00001 & short & NA\\ 
    Run H & delayed feedback & 0.0001 & long & NA\\ 
    Run I & delayed feedback & 0.00001 & long & NA\\ 
    Run J & delayed feedback & 0.01 & long & NA\\ 
    Run K & delayed feedback & 0.1 & long & NA\\ 
    Run L & delayed feedback & 0.00001 & long & solid body rotation\\ 
    Run M & delayed feedback & 0.00001 & long & Fraction of grav. rotation\\ \hline
  \end{tabular}
\end{center}
\label{t:sims}
\end{table}

An important diagnostic used to compare the different models was the mass
accretion rate across the inner boundary of the simulated grid.  This
was measured in ${\rm M}_\odot\,{\rm year}^{-1}$.  To calculate the mass flow,
the amount of mass in each cell on the inner boundary with a
negative (inward) velocity in the radial direction was summed.  This
is probably not completely accurate at the highest densities and mass
accretion rates as the inflow boundaries are not perfectly efficient
and sound wave may be reflected off the boundary.  However, this
should not change the results 
since these inaccuracies do not occur until the mass flow and density
reach physically unrealistic values.

\subsection{Radiative cooling}
\label{sec:cool}

The driving force behind the cooling flow is the thermal
bremsstrahlung and line radiation which removes thermal energy from
the ICM core.  This is modeled with the optically thin cooling law
\begin{eqnarray}
  \label{eq:rad}
  \Lambda = [C_1(k_B T)^\alpha& + C_2(k_B T)^\beta 
  + C_3] 0.704\,(\rho/m_p)^2 \nonumber\\
&\times 10^{-22}\mbox{ ergs}\mbox{ cm}^{-3} s^{-1},
\end{eqnarray}
which is the same law used by \citet{2002ApJ...581..223R} with the
coefficients $C_1=8.6\times10^{-3}$, $C_2=5.8\times10^{-2}$,
$C_3=6.4\times10^{-2}$, $\alpha=-1.7$, and
$\beta=0.5$ \citep{1993ApJS...88..253S}.  This cooling term is
proportional to $\rho^2$ which produces the eventual catastrophic
cooling as the center of the cluster becomes denser as it cools.
Below a minimum temperature (0.1\,keV), cooling was manually
truncated.  Even if we allowed material to cool below this limit, our
spatial resolution would be insufficient to follow the resulting
structures and additional physics not captured by eqn.~\ref{eq:rad}
would become applicable.

The cooling represents an extra term in the energy equation
(eqn.~\ref{eq:hd3}) and is implemented as an explicit source term.  As
with all physical processes, there is a maximum allowed time step for
numerical stability associated with the cooling, but in our case it is
always above the normal hydrodynamic CFL condition and need never be
implemented.

Due to the density dependence of the radiative cooling, it is only
relevant in the inner regions of the cluster.  This is important for
three reasons.  First, as the core cools, the inner regions loses
pressure support.  This causes the inward flow of material, some part
of which could fuel the accretion on the central compact object.
Second, as the inner regions become denser due to the sagging, the
cooling increases.  This causes the eventual runaway cooling whose
absence represents part of the cooling flow problem.  Finally, since
the outer region does not cool appreciably with a Hubble time or so,
it represents are large reservoir of hot material which could
potentially be exploited to balance cooling (although this does not
seem possible in purely hydrodynamic models).

\begin{figure*}
  \centering
  \epsscale{0.75}
\plotone{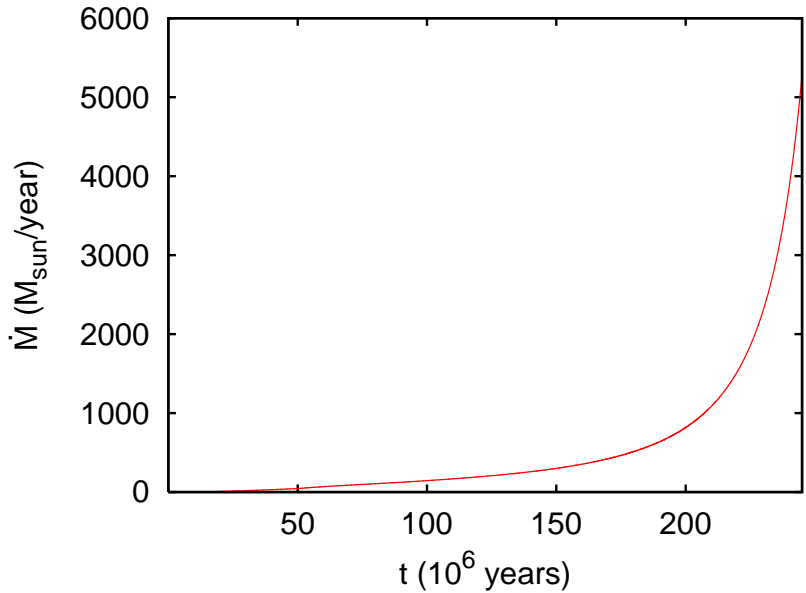}
  \caption{Mass accretion rate for pure cooling (Run A).}
  \label{fig:mdotA}
\end{figure*}

To establish a control, our first simulation (Run A) followed the pure
radiative collapse of our model ICM atmosphere (i.e., a spherically
symmetric homogeneous cooling flow).  The evolution of this system was
very simple.  The cluster cools, primarily in the inner regions.  As
it cools, the inner regions become denser.  As the gas becomes denser,
it cools quicker, and the process runs away (until numerical issues
with the very cool dense gas force us to terminate the simulation).
The mass accretion rate for pure cooling can be seen in
Figure~\ref{fig:mdotA}.  With no mechanism to stop or slow the
cooling, the mass accretion show a featureless, approximately
exponential increase. 
This unbounded cooling and mass accretion is our first example of a
catastrophic cooling.

\begin{figure*}
  \centering
  \epsscale{0.75}
\plotone{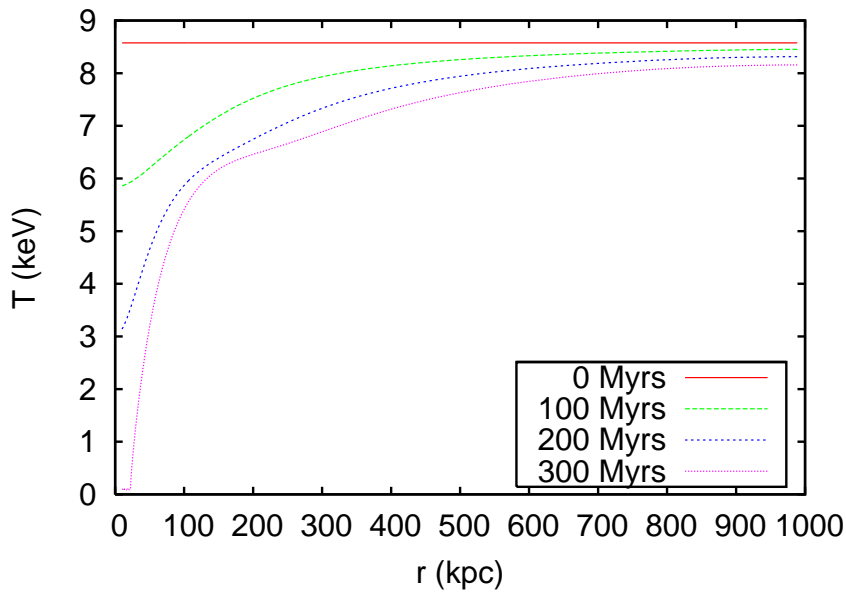}
  \caption{Radial temperature dependence for pure cooling (Run A).}
  \label{fig:rtA}
\end{figure*}

To diagnose the time taken for the system to undergo a cooling
catastrophe, we measure the time taken for each of our simulations to
exceed a mass accretion rate of $5000 {\rm M}_\odot {\rm yr}^{-1}$ (an
arbitrary ``large'' mass accretion rate).  For the pure cooling flow
run (Run A), this threshold mass accretion rate is crossed at
244.1 Myrs.  Material first fell below our imposed lower temperature
limit at 248.9 Myrs.  The radial dependence of temperature for Run A is
shown in Figure~\ref{fig:rtA}.  Initially the cluster is isothermal.
The inner regions cool first, while the outer regions barely change in
temperature.  As catastrophic cooling occurs, the temperature gradient
in the inner region becomes progressively steeper.

\subsection{Single AGN Outburst Models}
\label{sec:jet}

The onset of the cooling catastrophe in the homogeneous cooling flow
is well known and unsurprising.  Our hypothesis and the focus of this
work is the idea that jet activity by a central AGN can heat the ICM
core and prevent this catastrophic cooling.  As an initial exploration
of jet heating models, we follow the evolution and effects of a period
of jet activity in which the jet has a fixed and constant power for a
pre-defined duration.  In Run B, the jet is active for 50\,Myrs after
which it is completely shut off and the resulting ICM allowed to
evolve passively.  These cases are essentially 3-dimensional
generalizations of the axisymmetric simulations of
\citet{2002MNRAS.332..271R} and share many characteristics with (but are
higher resolution than) \citet{2003MNRAS.339..353B}.  The axisymmetric
simulations of \citet{2005A&A...429..399Z} provide another example of
this type of single burst jet heating.  This can be
viewed as a zeroth order approximation to AGN feedback.

In detail, the action of the AGN is modeled as a jet of low density
material injected from the inner radial edge of the simulation grid
using an inflow boundary condition.  The jet has a density of $1/100$
of the initial inner ICM density and (at the inner boundary) is in
pressure equilibrium with the initial inner ICM.  Therefore, the
internal sound speed of the injected jet material is 10 times that in
the initial ICM atmosphere.  The jet has an opening angle of
$15^\circ$.  In the case of a single jet outburst of Run B, a Mach
number of around $10.5$ compared to the background material was used
resulting in a total kinetic luminosity of $9.8\times 10^{45}\,{\rm
erg}\,{\rm s}^{-1}$.  We also modeled a rather weaker jet with kinetic
luminosity of $8.9\times 10^{45}\,{\rm erg}\,{\rm s}^{-1}$ in Run C
(which was otherwise identical to Run B).

\begin{figure*}
  \centering
  \epsscale{1}
\plotone{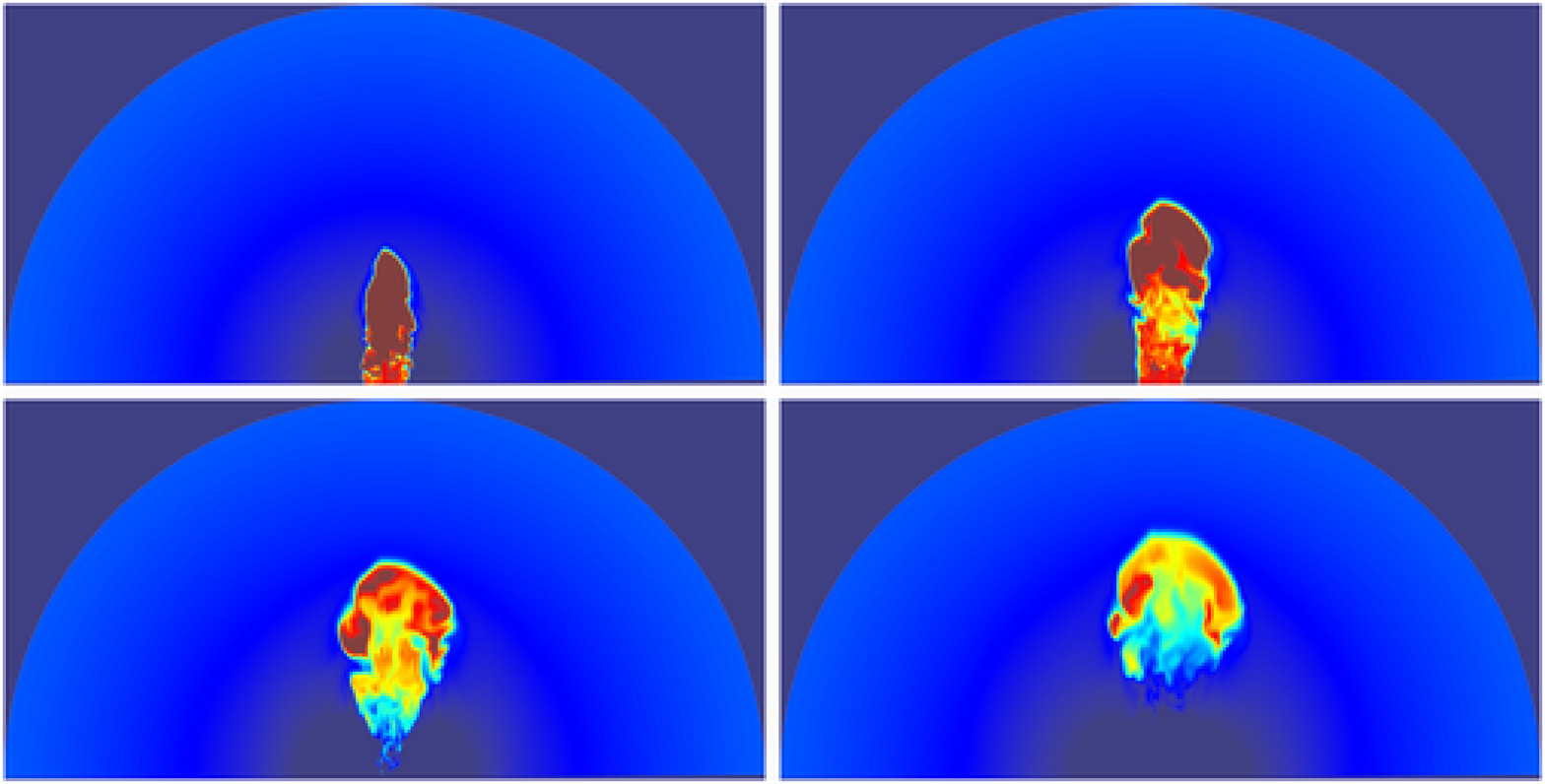}
  \caption{Entropy plots for single jet (Run B) at t=50, 125, 250, and
    375 Myrs.  The semi-circle in each plot represents the outer edge
    of the simulated grid at 1000 kpc.}
  \label{fig:entropy}
\end{figure*}

Plots of entropy are shown at various times for Run B in
Figure~\ref{fig:entropy}.  As discussed in \citet{2002MNRAS.332..271R},
plots of entropy are a good tool for differentiating shocked jet
material from background material.  Initially, the jet carves a path
through the ICM, and terminates in a shock.  It can be seen that the
jet channel is surrounded by a backflow of shocked (high entropy) jet
material.  This shocked material is over-pressurized compared to the
background, and expands laterally into a cocoon structure.  The cocoon
is separated from the background by a contact discontinuity.  
Rayleigh-Taylor (RT) and Kelvin-Helmholtz (KH) instabilities work 
to shred this cocoon and mix in background material.
After the jet has been turned
off, the cocoon is left to evolve passively.  Buoyancy forces cause it
to rise, leaving the cluster core.  Once the cocoon completely
detaches itself from the core, it takes on the appearance of a rising
bubble, which spreads out and fades somewhat into the cluster
background (much like observed ghost bubbles) until it leaves the
computational grid.

\begin{figure*}
  \centering
  \epsscale{0.75}
\plotone{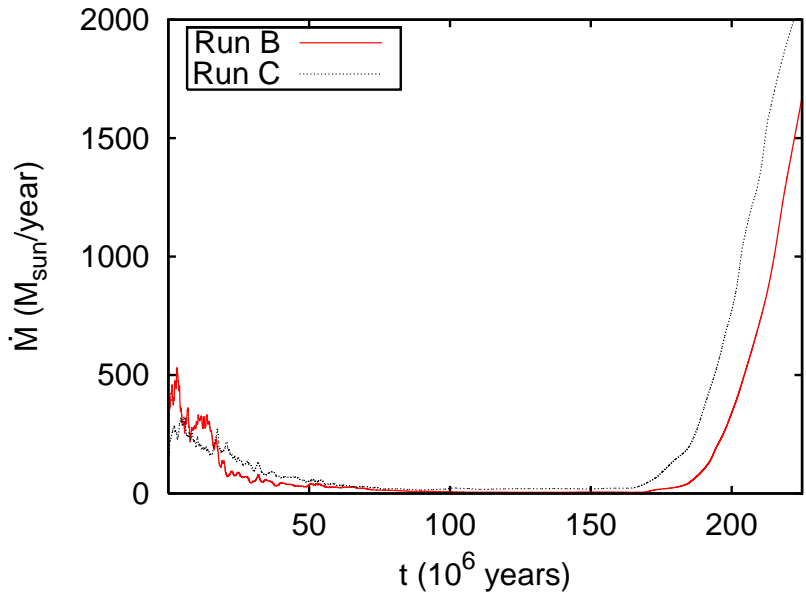}
  \caption{Mass accretion rate for single burst (Run B and Run C).}
  \label{fig:mdotBC}
\end{figure*}

Figure~\ref{fig:mdotBC} shows mass accretion for Run B and Run C.
Although these differs slightly in the timing, they qualitatively show
the same behavior.  Initially, there is moderate mass flow (a few
hundred ${\rm M}_\odot {\rm yr}^{-1}$).  Ironically, the jet-activity
is actually responsible for this initial enhanced period of inflow;
some matter in the innermost region of the ICM core becomes caught in
the backflow that results from the onset of the of the jet activity
and is swept across the inner boundary.  After the initial spike, the
mass accretion rate drops off to very low values while the jet is on
and remains very low for over 100 Myrs after the jet stops due to the
shock heating of the ICM core.  The cooling flow then starts to
re-establish and the mass flow begins to increase (eventually)
catastrophically as it does in the pure cooling case.  The time for
catastrophic cooling has been delayed somewhat from the pure cooling
case.  For Run B, $\dot{M}$ does not reach $5000\,{\rm M}_\odot {\rm yr}^{-1}$
until 306\,Myrs (more than 60\,Myrs later than the pure cooling model).
Material falls below the minimum cooling temperature at 305 Myrs
(i.e., at essentially the same time that the mass accretion rate
crosses our ``catastrophe'' threshold).  The weaker jet, Run C,
reaches the catastrophic point at 285 Myrs, which is only slightly
delayed from the pure cooling case.  The temperature falls below our
floor at 274 Myrs (also slightly later than the pure cooling case).

\subsection{Feedback Models}
\label{sec:feed}

The radio-loud AGN feedback hypothesis would argue that the central
AGN acts such as to maintain a long term balance of jet heating and
radiative cooling of the ICM.  Furthermore, this must occur in systems
with widely varying radiative luminosities.  Clearly, this mechanism
requires that the AGN power be somehow regulated by conditions within
the ICM so that it does not underheat or overheat the ICM core [see
\citet{1995MNRAS.276..663B} for early work on AGN feedback on cooling
flows].  In this work, we explore a set of models in which the kinetic
luminosity of the jets is connected to the mass accretion rate in the
cooling flow.  Unlike the other effects (cooling, jet propagation,
etc.), feedback is the only one that does not directly represent
fundamental physics of the cluster gas.  Instead, it tells us
something about fueling in the central engine of the AGN.  This
provides us with the most freedom in how to implement it.  Due to the
large range of scales involved, it is not possible to directly model
feedback.  This would require resolving gas for the outer edges of the
cluster (nearly a megaparsec) all the way down to the inner accretion
disc (parsec scales).  A direct assault on this problem would not be
practical at the current time even for an Adaptive Mesh (AMR)
code. Instead, we must use some simplified model for feedback.

In our first set of feedback simulations (Runs D and E), the jet is
injected with a kinetic energy proportional to the instantaneous mass
accretion rate across the inner radial boundary of our simulation
domain.  This is achieved by modulating the injection velocity to be
\begin{equation}
  \label{eq:vjet}
  v_{jet}=\left(\frac{2\eta\dot{M}c^2}{A\rho}\right)^{\frac{1}{3}},
\end{equation}
where $\dot{M}$ is the mass flow rate across the inner radial boundary
of the simulation, $c$ is the speed of light, $\rho$ is the density of
the injected jet material, $A$ is the area of the jet ``nozzle'' on
the inner boundary, and $\eta$ is the efficiency with which the rest
mass energy of the ICM cooling flow is converted into jet power.  The
appropriate choice for $\eta$ is far from clear.  In the extreme case
that {\it all} of the matter that flows across the inner boundary was
to accrete onto the central supermassive black hole, $\eta$ would be
the jet-production efficiency of the actual black hole accretion disk
itself.  For an efficient disk, we would have $\eta\sim 0.1$, although
higher efficiencies are possible if the black hole is
spinning \citep{1973NovikovThorne} or if magnetic coupling within the
radius of marginal stability becomes
important \citep{1999ApJ...522L..57G,2003MNRAS.341.1041A}.  However,
as has been extensively discussed, supermassive black holes in the
centers of giant elliptical galaxies accreting at a modest rate from
the hot interstellar medium may well have a significantly smaller
efficiency ($\eta\sim 10^{-3}$). Much of the remaining energy is
advected across the event horizon or driving a slow, uncollimated wind
from the disk.  Furthermore, it is plausible that only a small
fraction of the mass that flows across the inner boundary of our
simulation domain (located at 10\,kpc) actually enters the sphere of
influence of the massive black hole, with some of the remaining going
to fuel low-level star formation.

\begin{figure*}
  \centering
  \epsscale{0.75}
\plotone{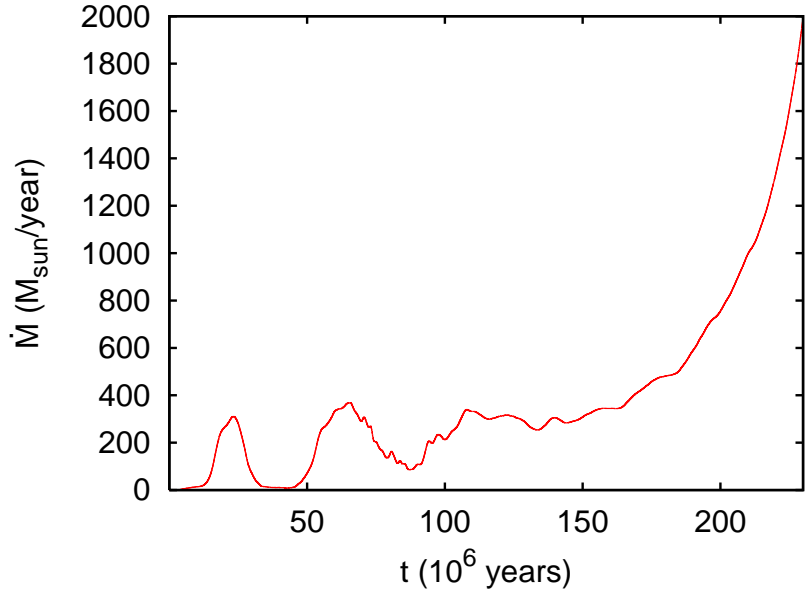}
  \caption{Mass accretion rate for feedback model (Run E).}
  \label{fig:mdotE}
\end{figure*}

Motivated by this discussion, we explored two cases of simple feedback
with efficiencies of $\eta=10^{-4}$ (Run D) and $\eta=10^{-5}$ (Run E).
The resulting mass accretion for Run E (lower efficiency) is given is
Figure~\ref{fig:mdotE}.  The higher efficiency (Run D) shows the same
pattern.  Initially, the mass accretion rate cycles up and down in
response to the jet power (partly due to the backflow noted above), as
we would expect if feedback works to stop cooling.  However, by 150
Myrs, this breaks down.  From that point on, the mass accretion rate
increases, and the increasingly powerful jet does nothing to stop it.
The time for catastrophic cooling and for material to fall below the
minimum cooling temperature is not significantly affected by this type
of feedback.  Both runs have very similar catastrophic cooling times
with 246.7 Myrs for Run D and 245.43 for Run E. This is 
essentially the same as the time for the pure cooling model.

\begin{figure*}
  \centering
  \epsscale{0.5}
\plotone{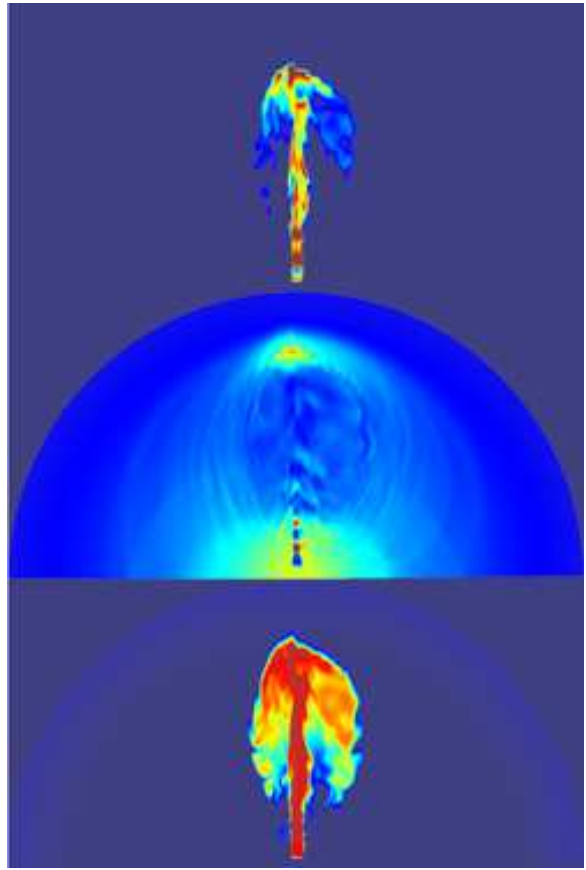}
  \caption{Temperature Map (top), Pressure Map (middle), Entropy Map
  (bottom) for immediate feedback jet.  The temperature map only shows
  the highest temperatures to pick out primarily jet
  material.  The thin, low density channel can be easily seen in
  temperature and pressure (and to a lesser extent entropy).  Only the
inner 254 kpc of the simulation is shown.}
  \label{fig:channel}
\end{figure*}

The reason for this failure of feedback to prevent the cooling
catastrophe can be seen in Figure~\ref{fig:channel}.  After the jet
has been active for a while (even at very low power) it clears a
low-density channel in the ICM.    As cooling and ICM accretion proceeds
along equatorial latitudes (i.e., in the plane perpendicular to the
jet axis), the increasingly powerful jet flows freely down the
pre-cleared channel.  This prevents the jet from depositing energy
near the cluster core.  Instead the kinetic energy is carried to the
head of the cocoon reasonably unimpeded.  With no energy deposition
near the core, cooling proceeds on the catastrophic course in the
equatorial regions almost as if there were no jet.

\subsection{Delayed Feedback Models}
\label{sec:delay}

The failure of the simple feedback model is, at least in part, related
to the formation of the low density channel along the jet axis.  We
note that \citet{2004MNRAS.348.1105O} also discuss the importance of
such channels within the context of their slow-jet simulations and
argue that the effective heating required the channels to fill in
between powerful outbursts.  Motivated by this, and in an attempt to
model more realistic feedback, we performed a set of simulations in
which a time delay was introduced between the mass accretion rate and
the response of the jet.  Clearly the immediate feedback of Run D and
E is not physically accurate.  Some time delay must be added to
account for the material travel time from the inner edge of the grid
to the accretion disk and onto the black hole.  We note that the time
for the relativistic jet to reach from the black hole and to enter the
computational domain is negligible.

In Runs F and G, a delay of 10 Myrs was introduced.  This is the sound
crossing time of the cluster center (the area from the inner edge to
$r=0$) and represents the minimum physically reasonable delay.  Runs
H, I, J, and K all have a delay time of 100 Myrs.  This is
approximately the dynamical time for the central galaxy and is a
reasonable ``best-guess'' at a physically plausible time delay.
Interestingly, this is also the approximate cooling time observed by
{\it Chandra} in the centralmost regions of cluster cooling cores.  
The efficiencies of these runs is listed in
Table~\ref{t:sims}.

\begin{figure*}
  \centering
  \epsscale{0.75}
\plotone{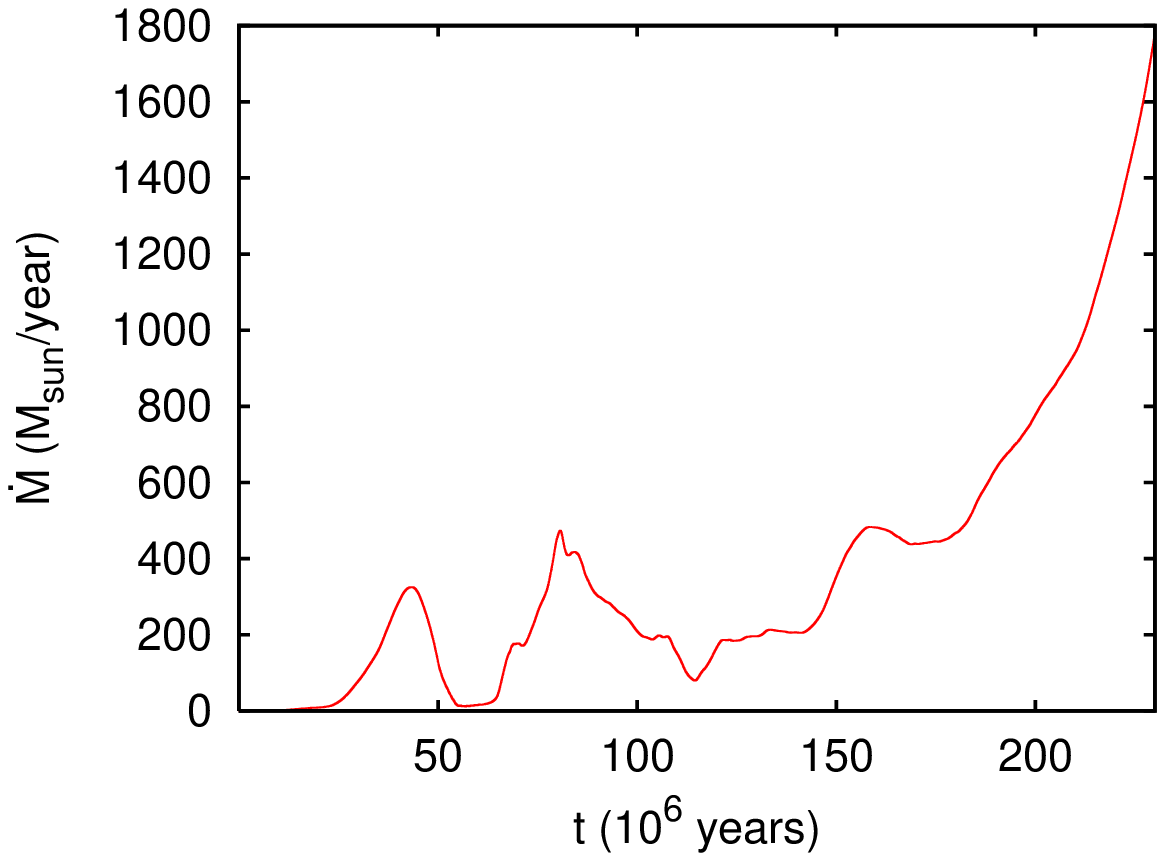}
  \caption{Mass accretion rate for low efficiency delayed feedback (Run G).}
  \label{fig:mdotG}
\end{figure*}

An example of the low efficiency delayed feedback with a short delay
can be seen in Figure~\ref{fig:mdotG}.  There is not much difference
between this and the immediate feedback.  There is also no essential
difference in the time to reach catastrophic cooling.  The only minor
effect we see here is that the lower efficiency run (G) has a slightly
lower rate of increase for the accretion, probably due to less
efficient channel formation.

\begin{figure*}
  \centering
  \epsscale{0.75}
\plotone{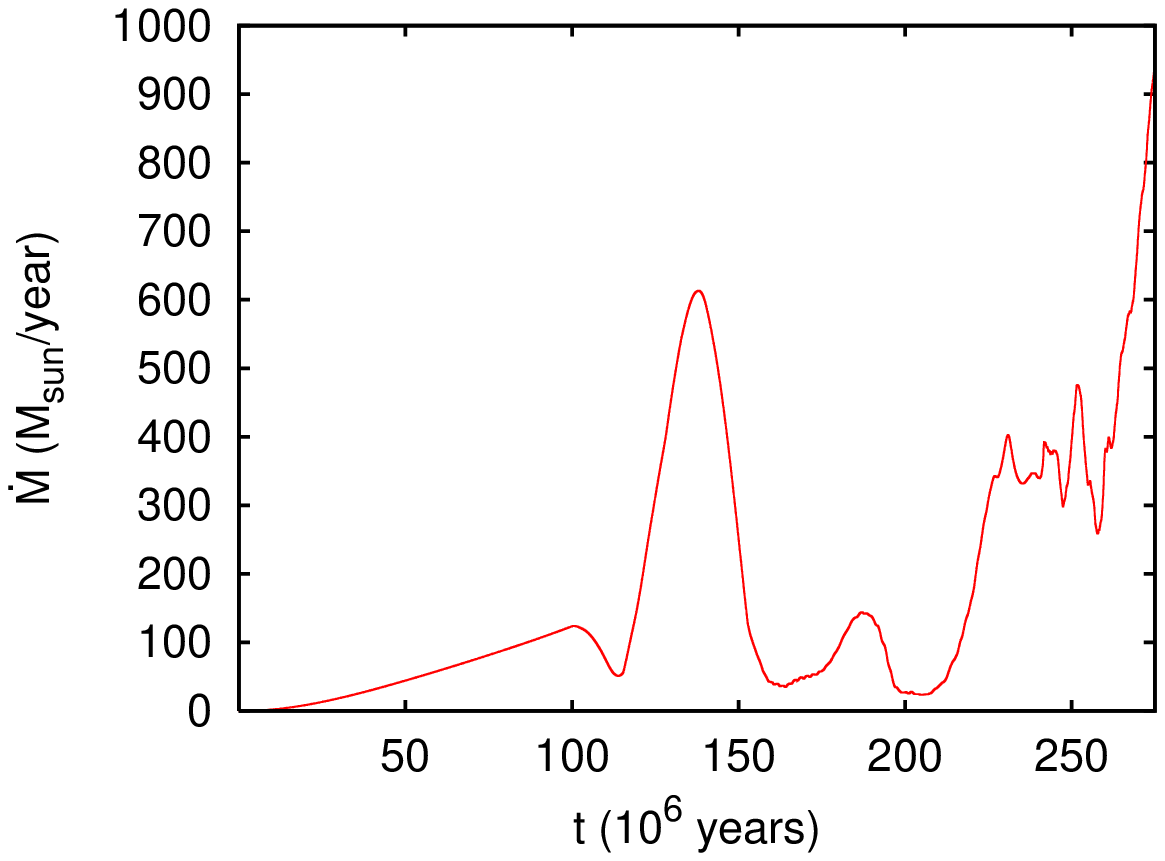}
  \caption{Mass accretion rate for low efficiency long delayed
    feedback (Run I).}
  \label{fig:mdotI}
\end{figure*}

For Runs H through K, the physically motivated delay of 100 Myrs was
used.  Runs H and I and essentially copies of Runs F and G with a
longer delay between the mass flow and the jet.  As
Figure~\ref{fig:mdotI} shows, this have a more significant effect on
the mass flow than the shorter delay had.  For the first 100 Myrs,
there was no jet, so the run proceeded as a pure cooling run.  The
onset of (weak) jet activity caused a sharp dip in the accretion,
followed by a spike (up to 600 $M_\odot {\rm year}^{-1}$) lasting for
about 50 Myrs (or half of the dynamic time).  However, after several
smaller smaller spikes and dips, runaway cooling proceeds in the
equatorial plane.  In these cases, the cooling catastrophe is delayed
as compared with the pure cooling model, but only by 40\,Myrs.  The
higher efficiency version (Run H) behaved similarly only it reached
the cooling catastrophe slightly faster (once again due to increased
channel formation).

Given the difficulties with the feedback models found above, we also
ran two simulations with {\it very} efficient jet production; Run J
and K possessed $\eta=0.01$ and $0.1$ respectively.  Run K was
performed in the spirit of an {\it extreme case} and we note that it
is not reasonable that accretion onto the central galaxy and then onto
the central AGN are both perfect, with the only loss coming from
relativity and the central black hole.  Even at an efficiency of 0.01
it is hard to come up with a plausible scenario for such nearly
perfect accretion.

\begin{figure*}
  \centering
  \epsscale{0.75}
\plotone{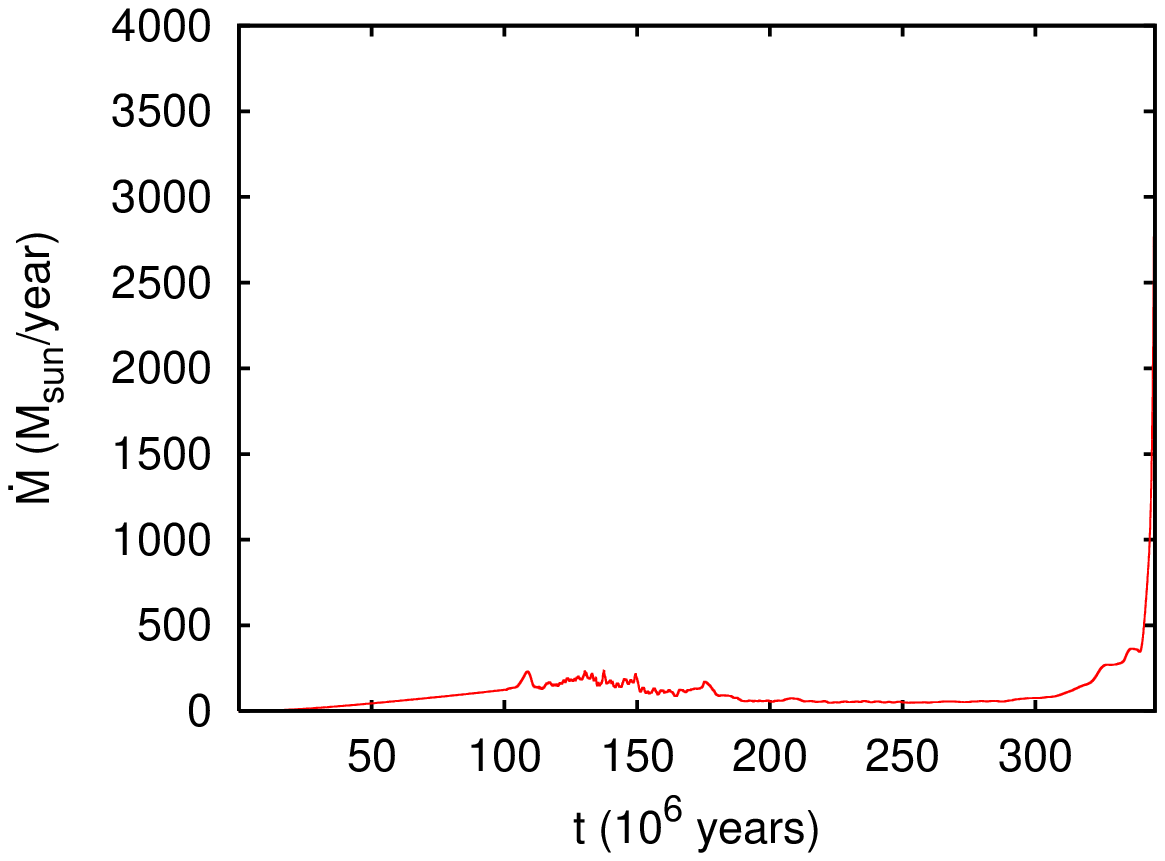}
  \caption{Mass accretion rate for very high efficient feedback (Run J).}
  \label{fig:mdotJ}
\end{figure*}

Even the extreme runs are not able to completely stop cooling.  For
Run J, the temperature falls below $T_{\rm min}$ at 245 Myrs.  This is even
sooner than in the pure cooling case.  Run K does not have this until
267 Myrs which is between the times for the other delayed runs and the
single jet runs.  The extreme efficiency feedback does, however,
significantly delay catastrophic cooling.  Run J does not experience
catastrophic cooling until 345 Myrs.  The mass accretion behavior is
shown if Figure~\ref{fig:mdotJ}.  Although it has taken longer for the
cooling to reach our catastrophic level, the final catastrophe is
extremely sharp, with the mass accretion rate exponentiating on
timescales of only 1\,Myr or so. This is much more dramatic than any
of the less efficient runs.  Run K holds of catastrophic cooling for
50 Myrs beyond Run J (398 Myrs), but it too becomes catastrophic
rapidly at that point.

\subsection{Feedback with Rotation}
\label{sec:rotation}

The AGN feedback and jet heating models described above clearly fail
to adequately heat the ICM core in a manner that stably offsets
radiative cooling.  Our final two simulations (Runs L and M) explore
the effect of an ICM atmosphere that is not initially static. The ICM
of real clusters is never, of course, a perfect hydrostatic
atmosphere.  Clusters have complex merger and formation histories that
leave an imprint on the dynamics of the
ICM \citep{1977ApJ...217...34B}.  In several
clusters observed with {\it Chandra}, the ghost cavities are not
symmetric about the cluster center (e.g., A4059, \citet{2002ApJ...569L..79H};
Perseus-A, \citet{2003MNRAS.344L..43F}), suggesting that they are buoyantly
rising in an ICM atmosphere which itself has velocity structure.
While a full exploration of this class of models is beyond the scope
of this paper, our final two simulations model the case of a rotating
ICM atmosphere.  Due to the assumed reflection symmetry in the plane
perpendicular to the jet axis, we are forced to consider only
rotations with an axis coincident with the jet axis.

In Run L, a delayed feedback model similar to Run I is examined in
which the initial ICM atmosphere is assumed to undergo solid body
rotation.  The rotation speed at the outer edge of the grid is set
equal to the sound speed of the cluster.  This leave the rotation in
the core at a very low value.  The results from this were nearly
indistinguishable (in cooling, not in detailed physical structure)
from Run I.  This is because the rotation in the core was not high
enough to mix the ICM effectively or stop accretion in any other way.
Without having gas moving supersonically in the outer parts of the
cluster (which is clearly unphysical), there is no way to get solid
body rotation to help offset cooling.

\begin{figure*}
  \centering
  \epsscale{0.75}
\plotone{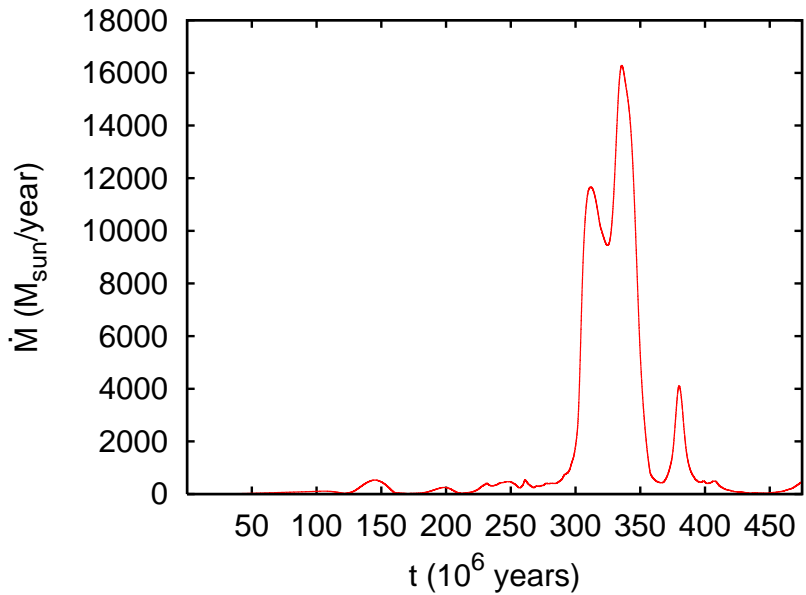}
  \caption{Mass accretion rate for rotating cluster with feedback (Run M).}
  \label{fig:mdotM}
\end{figure*}

In Run M a rotation law was chosen such that the centrifugal force
associated with the rotation is a fixed fraction (10\%) of the
gravitational force at that location.  This results in a rotation law
in which the angular velocity increases with decreasing radius until
one gets well within the ICM core, at which point the angular velocity
smoothly goes to zero at the center.  Figure~\ref{fig:mdotM} from Run
M shows the first accretion curve that differs significantly from the
previous patterns.  Initially, the accretion rate goes up and down in
response to the jet.  At around 240 Myrs, gas falls below the cooling
limit (similar time to the pure cooling and most feedback runs).  At
285 Myrs, the accretion passes the 5000 $M_\odot{\rm year}^{-1}$ limit and
continues to climb.  Unlike previous setups, it does not grow without
bound (or flatten out at some unrealistically high value).  Instead at
16,000 ${\rm M}_\odot{\rm yr}^{-1}$, the curve reverses and over the
course of 50 Myrs drops back down to very low levels (a few hundreds
of solar masses per year.  It then continues at that rate with only
small fluctuations for the duration of the simulation.

\begin{figure*}
  \centering
  \epsscale{1}
\plotone{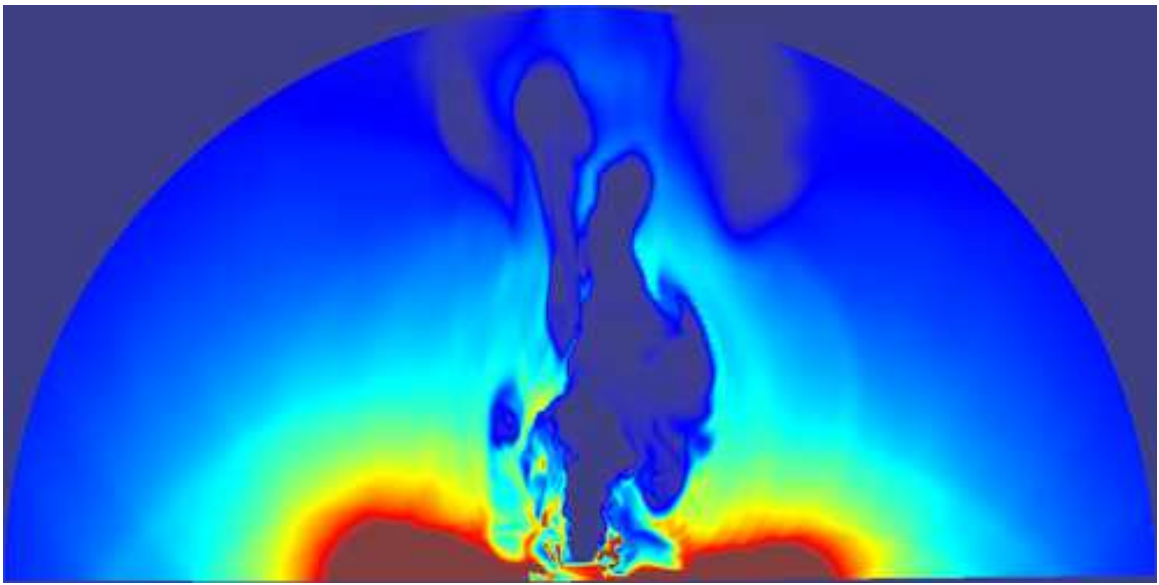}
  \caption{Dense cold disk formed in the inner regions of Run M.
    Density at 880 Myrs shown.  Only the inner 254 kpc of the
    simulation is shown.}
  \label{fig:disk}
\end{figure*}

In this case, the AGN behavior is actually rather incidental.  At
early times, the slowly rotating inner core of the cluster cools and
accretes.  Higher angular momentum material flows inwards and,
eventually, the ICM core becomes rotationally supported.  Given the
assumptions of our model, this material will conserve its angular
momentum and, irrespective of radiative cooling, will not accrete.
The main role of the AGN is to prevent accretion of ICM from the high
latitude regions within the centrifugal barrier.  Indeed, the cooling
catastrophe has {\it not} in fact been averted.  Rapid cooling into a
rotationally supported disk can clearly be seen in the late stages of
this simulation (Figure~\ref{fig:disk}).  Our mass accretion
diagnostic (based on mass flow
across the inner radial boundary of the simulation) fails to detect
this particular manifestation of the cooling catastrophe.

\section{Discussion}
\label{sec:disc}

Our work with simple hydrodynamic models show that it is more
difficult than previously assumed to halt a cooling flow through AGN
activity.  The ultimate hope (as in, for example, the one dimensional
models of \citet{2002ApJ...581..223R}) is that a proper coupling of
the AGN kinetic luminosity to the cluster gas will allow a long-term
balance to be established where the AGN heating balances the ICM
radiative cooling.  In our three dimensional simulations, we do not
see such balance.  Instead, the AGN jet seems to effectively clear a
low density channel through the ICM core, after which time the kinetic
energy of the jet is deposited well outside of the cooling core.
Thus, simulations with delayed feedback do not produce many multiple
outbursts.  Instead they produce one or two long outbursts with
varying jet power before the ICM core catastrophically cools.

Our modeled AGN feedback is not totally ineffectual at heating the
ICM.  Both the time for catastrophic cooling and the time for gas to
fall below our cooling limit are increased by the AGN in most cases.
The single jet cases are rather effective (delaying the catastrophe
for as long the most effective feedback simulations).  However, the
instantaneous feedback and the feedback with a time delay of the core
sound crossing time have basically no effect of the catastrophe time
(no more that a few Myrs).  The more realistic delays are capable of
delaying the cooling catastrophe by several tens of Myrs (more in the
unrealistically efficient cases).  The only exception to this
behavior was the rotating cluster modeled in Run M.  As was shown in
Figure~\ref{fig:mdotM}, there is a brief period of very high
accretion, following by an extended period of lower
accretion. However, as discussed above, catastrophic cooling is still
occurring into a rotationally supported disk.  The rotational support
prevents this gas from falling into the central region, but it is
still a large body of cool and cooling gas around the cluster center
which contradicts observations.

Due to the computationally expensive nature of our simulations and the
multiple parameters present, it was not possible to test every
permutation of the model parameters.  It is possible that there is
some preferred area of the parameter space where the AGN heating
balances the radiative cooling without producing unrealistic jets.
The potential existence of such a privileged area of the parameter
space does not change our conclusion as it would bring up a serious
fine tuning problem.  Cooling and AGN occur in a wide range of
clusters of varying masses and temperature.  For a regulatory
mechanism to work, it must be general enough to work in the different
clusters.  We believe that our simulations sample physically plausible
parts of parameter space, thus the failure to model successful AGN
feedback is an important result.

As discussed in the Introduction, however, there is significant
observational support for the notion that radio-galaxy heating can
provide a long-term balance against radiative cooling.  Thus, it seems
clear that our models are not capturing some ingredients which are
important in real systems.  In the rest of this section, we explore
some of these missing ingredients.

We must acknowledge that while we do model the jet, we may
well not capture all of the relevant jet dynamics.  Real AGN jets have
internal dynamics that can only be captured by considering
special-relativistic \citep{1997ApJ...479..151M} and
magnetohydrodynamics effects
\citep{2004MNRAS.348.1105O,2004MNRAS.350L..13O}.  Furthermore, our
simulation probably has insufficient resolution to capture the full
internal dynamics of a hypersonic jet even within the context of ideal
non-relativistic hydrodynamic.  These effects may allow the jet
momentum to be spread over a larger working surface (i.e., ``dentist
drill effect''), thereby helping alleviate the problems associated
with the jet drilling out of the ICM core.  The fact that our
simulations produce structures (cocoons and ICM cavities) with aspect
ratios comparable to those observed, however, suggests that this is
not a major deficiency of our simulations.

Before leaving issues associated with the properties of the jet
itself, we note that \citet{2004MNRAS.348.1105O}
and \citet{2004MNRAS.350L..13O} simulate the effects of slow ($\sim
10,000\,{\rm km}\,{\rm s}^{-1}$), heavier jets on a cooling ICM core.
In agreement with our conclusions, they also identify the importance
of channel formation and the need for the channels to collapse between
AGN outbursts.  A new phenomena that appears to emerge from these
simulations is the excitation of large-scale $g$-modes in the ICM
which persist (and slowly dissipate their energy in to the ICM) long
after the AGN activity has shut-down.  Tentative observational
evidence for $g$-modes may exist in the cluster surrounding the
radio-galaxy 3C~401 \citep{2005MNRAS.357..381R}.

Any aspect of the system which alleviates the problem of channel
formation will be potentially important in ensuring the jet energy is
thermalized in the ICM core.  The initial cluster in our simulations
start with a smooth, hydrostatic atmosphere.  Real clusters have
complex merger histories and turbulent ambient motions.  This
turbulence could help to distribute the energy from the jet more
uniformly, or to oppose the formation of channels.  Since the jets are
powerful and higher velocity than even mildly supersonic ambient
motions for most of their lifetime, it is uncertain how effective
turbulence would be in preventing channels and distributing energy.
More importantly, real AGN jets may precess over significant angles
(for a recent claim of jet precession in Perseus-A, see
\citet{2006MNRAS.tmp..150D}).  ICM channels may become irrelevant and
the kinetic energy thermalized in the ICM core if the jets can indeed
precess over large angles on a timescale comparable to or less than
the inner ICM cooling time.   

Finally, our results might be pointing to the importance of physics
beyond ideal hydrodynamics.  A very real possibility is that plasma
transport processes may be crucial.  Thermal conduction provides a way
to move thermal energy across a temperature gradient in the ICM and
may be essential in helping the system avoid the cooling catastrophe.
There is a large heat reservoir in the outer parts of the cluster
which could be exploited to warm the core.  Furthermore, thermal
conduction and viscosity could be important mechanisms for
thermalizing the sound waves and weak shocks driven into the ICM by
the radio-galaxy activity.  All of these processes may delay the ICM
cooling sufficiently to allow the jet-blown channels to refill, thus
allowing a long term balance to be established.  We note that there is
growing observational evidence for the action of thermal conduction in
real clusters (e.g.,
Perseus-A; \citet{2005MNRAS.360L..20F,2005MNRAS.363..891F}).  Magnetic
fields are also certainly present in the ICM and, at some level, an
MHD (or even a kinetic theory) treatment may be necessary.  This will
have an important effect on the stability of the buoyantly rising
bubbles and the mixing of the high entropy radio plasma with the
thermal ICM.  But magnetic fields may have more profound consequences.
Thermal conduction and plasma viscosity will both become very
anisotropic due to the suppression of transport processes
perpendicular to magnetic field lines, leading to qualitatively new
fluid instabilities [the magnetothermal
instability \citet{2000ApJ...534..420B} and the fire-hose
instability].  Cosmic ray pressure may also be relevant to driving ICM
convection \citep{2004ApJ...616..169C}.  In a cooling cluster core
with an embedded AGN, one can speculate that these instabilities drive
ICM turbulence which may, itself, be a crucial ingredient in the
dynamics of the system.  Of course, this would be above and beyond any
turbulence or bulk motions due to the merging history of the cluster.

An important consequence of this work is that it firmly underscores
the inadequacy of AGN feedback simulations that model AGN feedback
through isotropically-inflated bubbles at prescribed locations in the
ICM.  Models that involve pre-inflated bubbles, or injecting energy
isotropically at given locations in the ICM, will fail to capture
precisely that aspect of the physics that turn out to be crucial to
the failure of our models --- the development of a low density channel
along which the AGN outflow can travel unimpeded thereby carrying its
kinetic energy out of the cooling core.  For this reason, it is vital
that jet dynamics be included if a model is to properly address
AGN-halted cooling in a cluster core.

\section{Conclusion}

We have performed a set of high-resolution three dimensional
simulations of jetted-AGN embedded in the cooling ICM cores of galaxy
clusters in which the AGN power reacts in response to the cooling of
the cluster gas.  We make the first attempt to model regulated AGN
feedback in the hope of achieving a long-term balance between
radiative cooling and jet heating.  However, we fail to construct a
model in which AGN heating comes into long-term balance with the
radiative cooling.  The early time jet activity is extremely effective
at clearing a channel through the ICM core.  The jet at later times
flows freely down this channel and hence deposits its kinetic energy
well outside the region which is starting to undergo catastrophic
cooling.  This essential behavior is robust to changing the efficiency
or time delay characterizing the AGN feedback.  Hence, we argue that
some vital ingredient of AGN feedback is missing from our model; this
may include more realistic jets, ambient ICM motions/turbulence,
jet-precession, or additional ICM microphysics.Our results also
highlight the absolute necessity to follow the jet dynamics in any
attempt to address radio-galaxy heating of the ICM.

\section{Acknowledgments}

We thank Barry McKernan, Sebastian Heinz, Andy Young, and Derek Richardson for
extensive discussions throughout the course of this work.  We also
thank the anonymous referee as well as Brian McNamara, Carlos Frenk,
and James Binney for extremely useful comments on the original
manuscript.  We are grateful to the original developers of ZEUS-MP and
NCSA for providing the initial code to work with.  All simulations
reported in this paper were performed on the Beowulf cluster (``The
Borg'') supported by the Center for Theory and Computation (CTC) in
the Department of Astronomy, University of Maryland, College Park.  We
gratefully acknowledge support from Cycle-5 Chandra Theory and
Modeling Program under grant TM4-5007X and the National Science
Foundation under grant AST0205990.

\begin{appendix}
\section{ZEUS-MP}
\label{append:zeus}

All simulations in this work were done using a modified version of the
ZEUS-MP code.  Our version is based on the initial NCSA released
version (1.0b).  Our modifications, documentation, and several
supporting scripts have been made publicly available at
http://www.astro.umd.edu/$\sim$vernaleo/zeusmp.html under the same
terms as previous ZEUS releases.

Here, we describe our changes and modifications to ZEUS-MP.  First, the
code had to be ported to compile with a current FORTRAN 77 
compiler.  As different compilers implement the standard (and
the extensions to it) differently, we had to choose certain compilers
as our targets.  The Intel Compiler was used for the speed of the
executables it produces for x86 compatible machines.  For the sake of
portability, we also maintain compatibility with the GNU compilers.
None of these changes modify the behavior of the code.  Primarily this
involved removing multiply defined variables and cleaning up the
namelist routines and the namelists themselves.  Also, all filenames
that differ only by case (common in the FORTRAN 77 build process) were
changed to allow building on non-case sensitive filesystems.

To allow for long runs, the restart routines (which did not work for
parallel simulations) were completely replaced.  The new restart
routines work for parallel simulations and write to alternating
files to save disk space.  A wrapper script written in Perl is
provided to correctly pick the most recent restart dump (if present)
and start ZEUS-MP using that dump.  The wrapper script also performs
some basics checks of the integrity of the restart dumps before using
them as incomplete or missing restart dumps have proven far more
likely in parallel simulations than in single processor simulations.

Despite the fact that ZEUS is written nearly entirely in FORTRAN 77,
the C preprocessor is used heavily to allow for conditional inclusion
of code.  Some portions of this preprocessor code needed to be
fixed;  many preprocessor directives were not properly matched up or
where improperly nested resulting in preprocessor flags that only
worked for one of the two possible values.  Also, some preprocessor
directives did not enclose all of the code relevant to a given
option.  This made it impossible to completely turn certain options
off in the code.
The post-processor was modified to be useful for large
number of output files and will eventually be completely replaced
with a more general post-processor.  

Several new problem specific routines were added.  A routine to update
boundary values during a run was added.  A number of other changes and
bug-fixes were also made, primarily involving geometry specific bugs.
Also, the build process was updated, relying on improved makefiles
and a custom Perl script.  

To insure that ZEUS-MP is portable, we have tested and run benchmarks
on several different systems.  We are aided by the fact that ZEUS uses
NCSA's hdf4 format as its primary output format.  By using a portable
output format, we do not have to worry about endian issues with our
output files.  The one place we break this is in the restart dumps which are
produced as unformatted FORTRAN binary data.  We
have run and used ZEUS-MP on a variety of GNU/Linux distributions
(both 2.4.x and 2.6.x kernels) on both workstations and a Beowulf cluster.
AMD Athlon processors, AMD Opteron processors (in 32 bit mode only as
that is all we currently have available), and Intel Pentium 4
processors have primarily been used.  We have also compiled and run
ZEUS-MP on Apple's Mac OS X (darwin) on the G4 PowerPC processor.  We
have also attempted to run ZEUS-MP on a Sun Ultra80 (UltraSPARC II
processor) running Solaris 8.0.  As we were not able to successfully
use the MPI library needed, we were not able to run ZEUS-MP.  This
is the one weakness in our portability.  If a working MPI library
(preferably lam-mpi) and the hdf4 library cannot be compiled, ZEUS-MP
cannot run on a system.  This should not be a problem on any modern,
Unix-like system.

\subsection{Performance}

\begin{figure*}
  \centering
  \epsscale{0.75}
\plotone{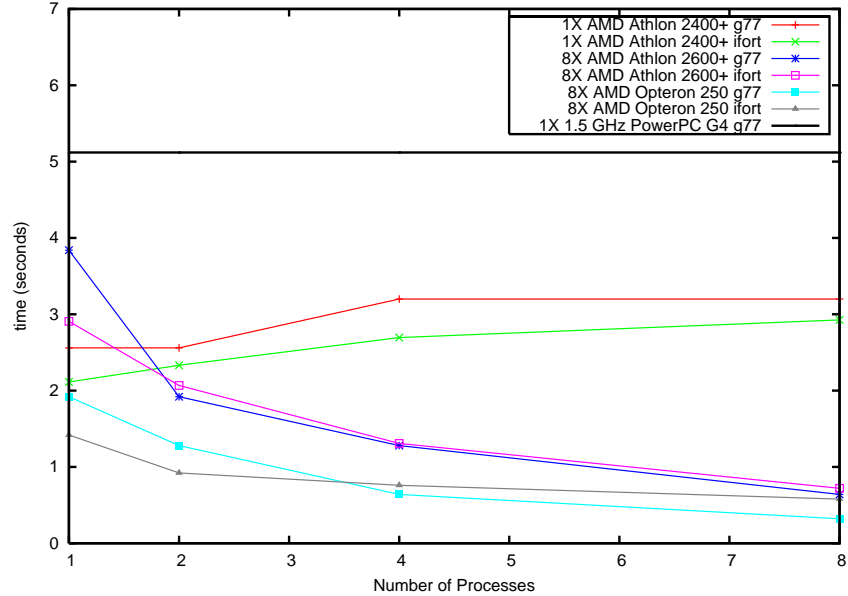}
  \caption{ZEUS-MP run time benchmarks for a variety of processors on
    a $100^3$ cell grid.}
  \label{fig:benc}
\end{figure*}

To test performance and scaling, a simulation was run with a $100^3$
Cartesian grid for 200 time steps in pure hydrodynamics mode with
radiative cooling added.  This simulation was run as one to eight
processes (not necessarily processors).  For single processor
machines, this means 
multiple processes on one processor (and presumably a large
performance penalty).  For the cluster tests this is not a problem.  The
timing results are shown in Figure~\ref{fig:benc}.  There are two
surprises in these results.  The first is that at 8 processors, code
compiled with the GNU compilers seems to outperform the code from the
Intel compiler (when used on AMD processors).  This is surprising as
the Intel compiler is widely considered to produce the fastest code
for x86 processors.  The second surprising feature is that the G4,
while it performs far below anything else in the tests, does not
appear to have any penalty for multiple processes on a single processor.

\begin{figure*}
  \centering
  \epsscale{0.75}
\plotone{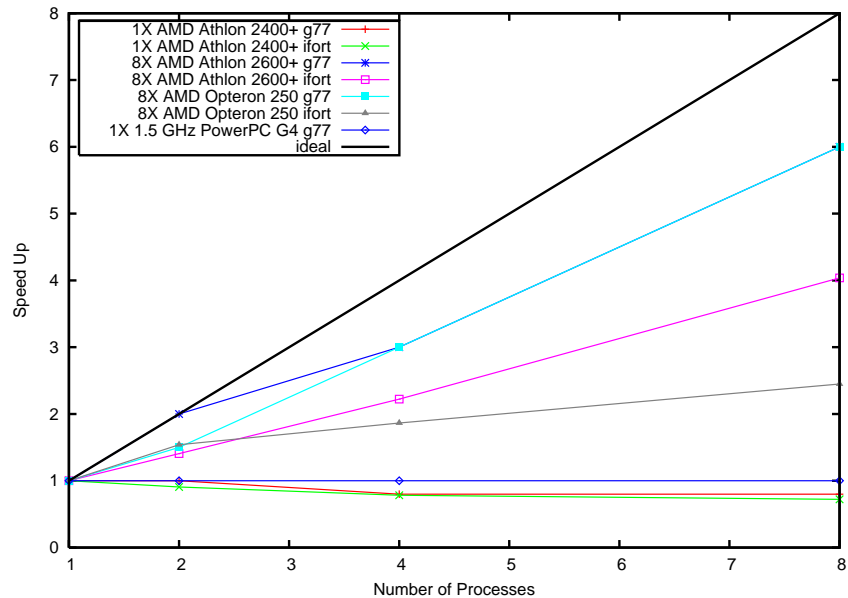}
  \caption{ZEUS-MP speed up tests for a variety of processors on
    a $100^3$ cell grid along with the theoretical performance.}
  \label{fig:speedup}
\end{figure*}

The results from the scaling test are shown in Figure~\ref{fig:speedup}.  While
at no point beyond two processors do we get the theoretical linear
increase, we continue to get a decent speedup (with a constant slope
in most cases) up to at least 8 processors, with the code produced
with the GNU compiler getting much closer to linear than the Intel
compiler.

\end{appendix}

\bibliographystyle{apj}
\bibliography{paper05}

\end{document}